\documentclass[10pt,a4paper,epsf]{article}
\usepackage{cite}
\usepackage[titletoc,toc,title]{appendix}
\usepackage{amsmath,amsfonts}
\usepackage{graphicx}

\DeclareMathOperator{\Tr}{Tr}
\DeclareMathOperator{\sech}{sech}

\def\be{\begin{equation}}   \def\ee{\end{equation}}

\begin{document}

\title{One-Loop Quantum Correction to the Mass of the Supersymmetric Kink in (1+1) Dimensions Using the Exact Spectra and the Phase Shifts}
\author {F. Charmchi\footnote{E-mail: f$\_$charmchi@sbu.ac.ir},
S.S. Gousheh\footnote{E-mail: ss-gousheh@sbu.ac.ir}
and S.M. Hosseini\footnote{E-mail: Morteza\_Hosseini@sbu.ac.ir}\\
{\small{\it Department of Physics, Shahid Beheshti University G.C., Evin, Tehran
19839, Iran}}}
\maketitle
\begin{abstract}
We compute the quantum correction to the mass of the kink at the one-loop level in (1+1) dimensions with minimal supersymmetry.
In this paper we discuss this issue from the Casimir energy perspective using phase shifts
along with the mode number cut-off regularization method.
Exact solutions and in particular an exact expression for the phase shifts
are already available for the bosonic sector.
In this paper we derive analogous exact results for the fermionic sector.
Most importantly, we derive a unique and exact expression for the fermionic
phase shift, using the exact solutions for the continuum parts of the spectrum
and a prescription we had introduced earlier.
We use the strong and weak forms of the Levinson theorem merely
for checking the consistency of our phase shifts and results,
and not as an integral part of our procedure.
Moreover, we find that the properties of the fermionic spectrum, including bound and continuum states,
are independent of the magnitude of the Yukawa coupling constant $\lambda$, and that the dynamical
mass generation occurs at the tree level.
These are all due to SUSY and are in sharp contrast to analogous models without SUSY, such as the Jackiw-Rebbi model,
where $\lambda$ is a free parameter.
We use the renormalized perturbation theory and find the counterterm which is consistent with supersymmetry.
We show that this procedure is sufficient to obtain the accepted value for the
one-loop quantum correction to the mass of the SUSY kink which is $-\frac{m}{2\pi}$.
\end{abstract}

\newpage
\tableofcontents

\newpage

\section{Introduction}
\label{sec:intro}
\setcounter{equation}{0}

Since early 1970\rq{}s the calculation of the quantum corrections
to the mass of the solitons in (1+1)-dimensional theories have received a great deal of attention.
In 1974, Dashen et al.\cite{Dashen} calculated the leading quantum correction to the bosonic
kink mass in $\phi^4$ theory.
Therein, the authors put the system in a box of length $L$ and
imposed periodic boundary conditions on the meson field leading to discretized frequencies,
and then used the mode number cut-off for the ultraviolet divergences.
They added a logarithmically divergent mass counterterm whose finite part was fixed by setting the tadpoles equal to zero in the trivial
background.
These soliton models were later extended to include supersymmetry (SUSY).
The issue of quantum corrections to the mass of soliton in (1+1)-dimensional supersymmetric theories has a long and controversial history.
Initially it was thought that in the computation of the quantum corrections to the mass of the supersymmetric solitons,
the bosonic and fermionic contributions, exactly cancel.
In that case, the Bogomolnyi-Prasad-Sommerfield (BPS) saturation condition would be preserved at the quantum level.
The BPS bound ensures that the soliton mass $(M)$  is greater than or equal to the magnitude of the central charge ($Z$)
to all orders in the quantum theory.
At the classical level this bound is saturated, i.e. $M=Z$.
In 1979, Schonfeld\cite{Schonfeld1979} computed the one-loop radiative correction to the mass of the SUSY kink using
the supersymmetry algebra along with the mode number cut-off regularization method.
There, a nonzero negative quantum correction, i.e. $-\frac{m}{2\pi}$, to the SUSY kink mass in (1+1) dimensions has been obtained.
Later, Kaul and Rajaraman\cite{Kaul} showed that the energies of solitons in the presence of fermionic fluctuations
receive a nonzero, finite but positive quantum correction.
It was believed that if one analogously repeats the same procedure for SUSY kink mass as it has been used for the bosonic kink mass \cite{Dashen},
no exact result can be reached since the quantum mass of the kink in SUSY theories
depends both on the choice of the boundary conditions and the regularization methods\cite{Nastase1999,SVV1999}.
In ref.\cite{Rebhan1997} it was claimed that
the Casimir energy, and therefore the soliton mass,
depends on the right choice of the cut-off and only mode number cut-off yields correct results.
However, later in 1999, Graham and Jaffe \cite{Graham1999} resolved the subtleties regarding cut-offs, boundary conditions,
and the counting of states that had plagued earlier calculations.
Therein, the authors correctly obtained the one-loop corrections to the energy and central charge of the supersymmetric kink.
The latter result is necessary to understand why the energy correction does not violate the BPS bound.
In their strategy they used the continuum density of states to write the Casimir energy in terms of the phase shifts and
regularized the continuum integral by subtracting the divergent Born approximations from the density of states. For this purpose, they expanded the scattering phase shifts as a Born series in the strength of the potential.
They also used the Levinson theorem as an integral part of
their procedure\cite{Graham1999,Jaffe1999,Graham1,E.Farhi2000}.
Later in 2000, Litvintsev and Nieuwenhuizen\cite{Litvintsev2000}
presented a generalized momentum cut-off regularization scheme, and
obtained the correct result for the SUSY kink mass.

Since then, many other authors have used several different approaches to compute the one-loop quantum correction
to the mass of the kink in (1+1) dimensions with minimal supersymmetry (${\cal N}=1$).
The most important of which are as follows:
First, computation of the Casimir energy using the mode number cut-off
\cite{Rebhan1997,Nastase1999,Raj1996,Kaul}.
The second one is an approach based on the existence of the anomaly in the expression for the central charge\cite{SVV1999}.
They used dimensional regularization by higher derivatives to calculate the anomaly in the central charge and
showed that there is an identical correction
to the central charge, so that the BPS saturation condition does not disappear at the quantum level to all orders in the weak coupling regime.
They used SUSY boundary conditions.
In their work they separated the soliton from boundaries to find the correct value of the SUSY kink mass.
As expressed in that article, they neglected \lq\lq{}the boundaries of the fiducial interval.\rq\rq{}
The third is using the heat kernel and the zeta-function analytic continuation method\cite{Bordag}.
In the literature authors now agree on the correct value for the correction to the SUSY kink mass in (1+1) dimensions
\cite{Goldhaber1,Goldhaber2,Goldhaber3,Litvintsev,Litvintsev2000}, i.e. $m^{(1)}_{\textmd{susy}}=- \frac{m}{2 \pi}\, .$

We believe that there are still some intricacies in this problem that have not been fully explored.
In this paper we present an unambiguous and robust procedure to find the exact
quantum correction to the SUSY kink mass in (1+1) dimensions.
We make the usual assumption that the back-reactions of both the boson and fermion fluctuations on the kink is negligible\cite{Jackiw1,Leila}.
Our procedure is based on the mode number cut-off along with the use of the exact forms of the phase shifts and the counterterm which
is consistent with SUSY.
For this purpose we need the exact solutions of the coupled boson-kink as well as the fermion-kink systems.
The former is already available and in this paper we shall derive and present the latter, which to the best of our knowledge has heretofore been absent.
The exact fermion-soliton solutions include the bound states and their energies, and the continuum states
whose asymptotic forms provide us with the information necessary for the computation of the phase shifts.
We investigate the difficulty with defining the phase shift
for the spinor $\psi(x,t)$ when solitons are present,
since the upper and lower components in general have different phase shifts.
Our prescription to overcome this problem is to define the phase shift to be the average of those two phase shifts\cite{Gousheh2}.
We check the consistency of the resulting fermionic phase shifts by using
the strong form of the Levinson theorem\cite{Gousheh2}, as well as its weak form\cite{Levinson}.
We have already used this prescription in several different models with consistent results\cite{Gousheh2,Gousheh3}.
We have checked that the fermionic phase shift obtained in this problem
by our general prescription matches the one obtained in\cite{Graham1999} using a different
procedure and tailored for cases where the potentials appearing in the decoupled
equations for the components of the spinor are proportional to $\sech^{2}(x)$.
We have been careful to properly identify and retain all of
the finite parts in the calculation of the Casimir energy, since they are crucial for the calculation of the correction to the SUSY kink mass.
We should mention that in our approach, there is no ambiguity in the choice of boundary conditions,
since we obtain the one-loop quantum correction to the SUSY kink mass
using the exact forms of the bosonic and fermionic phase shifts.
We like to emphasize the fact that we have used the Levinson theorem
merely as a checking utility, and not as an integral part of our procedure.
That is, using our procedure there is no need to use any fine-tunning such as
the one provided by the Levinson theorem.
Our analysis reveals a dynamically generated mass for the fermion at the tree level,
and we explain some of its intricacies including its discontinuous nature, as mandated by SUSY.

This paper is organized in six sections: In section \ref{sec:minimal} we begin with some basic definitions of minimal supersymmetry.
We write the Lagrangian describing our supersymmetric model at the end of this section.
In section \ref{sec:boson} we find the kink solution and exhibit the bosonic quantum fluctuations in the kink sector.
They include two bound states and a set of continuum states.
In section \ref{sec:fermion} we derive the coupled fermion-kink field equations and show that the exact solutions are
in the form of Gauss's hypergeometric function.
We derive the exact eigenvalues and eigenfunctions for these equations and compute
the phase shifts of the fermionic fluctuations.
Also, we check the consistency of the resulting phase shift with the Levinson theorem.
In section \ref{sec:one-loop} we use the exact expressions for bosonic and fermionic spectrum
along with the expression for the bosonic and fermionic contributions to the SUSY counterterm
in order to compute the one-loop
quantum correction to the SUSY kink mass in (1+1) dimensions.
Finally, in section \ref{sec:conclusion} we summarize and discuss our results.

\section{The minimal SUSY in (1+1) dimensions}
\label{sec:minimal}
\setcounter{equation}{0}

We start with the definition of superspace in (1+1) dimensions
which we use to work with the supesymmetric theories\footnote{Our notation is the same as that in \cite{SVV1999} and that in \cite{Litvintsev}.}.
We construct the superspace by adding a two-component real Grassmann variable
$\theta_\alpha =\{\theta_1, \theta_2\}$ to the two-dimensional
space-time $x^\mu = \{t,x\}$.
The following coordinate transformations subjoin SUSY to the Poincare transformations
\cite{SVV1999,Aitchison}:
\be
x^\mu
\to x^\mu  - i\bar\theta\gamma^\mu \varepsilon\, ,
\quad \theta_\alpha \to \theta_\alpha +\varepsilon_\alpha\, ,
\label{str}
\ee
where $\bar\theta=\theta\gamma^0$ and $\varepsilon$ is some constant Grassmannian spinor.
We shall use $\gamma^0 = \sigma_2$ 
and $\gamma^1 = i\sigma_3$  as a representation for the $\gamma$ matrices in two dimensions.
The real superfield $\Phi(x,\theta)$ is a function defined in superspace
and has the form
\be
\Phi(x,\theta)=\phi(x) +\bar\theta\psi (x)
+\frac{1}{2}\bar\theta\theta F (x)\, ,
\label{superfield}
\ee
Where $\theta$ and $\psi$ are real two-component spinors.
It is straightforward to show that the superspace transformations (\ref{str})
induce the following SUSY transformations:
\be
\delta\phi = \bar\varepsilon \psi\,
,\quad \delta\psi=-i \, \partial_\mu \phi \,\gamma^\mu\varepsilon
+ F \varepsilon\, ,
\quad \delta F=-i \bar\varepsilon \gamma^\mu \partial_\mu \psi \, .
\label{susytr}
\ee
These are the minimal, ${\cal N}=1$, supersymmetry transformations in (1+1) dimensions.

The associated action which is invariant under the SUSY transformations is
\be
S=i\int {\rm d}^2\theta\, {\rm d}^2x \left\{ \frac{1}{4}\bar D_\alpha
\Phi
D_\alpha\Phi +{\cal W}(\Phi)\right\}\, ,
\label{action}
\ee
where ${\cal W}(\Phi )$ is a superpotential. Here the spinoral derivatives are given by
\be
D_\alpha=\frac{\partial}{\partial\bar\theta_\alpha}
- i (\gamma^\mu\theta)_\alpha\partial_\mu\, ,
\quad {\bar D}_\alpha=-\frac{\partial}{\partial\theta_\alpha}
+ i (\bar\theta\gamma^\mu )_\alpha\partial_\mu\, ,
\ee
so that
$$
\left\{D_\alpha,\bar D_\beta\right\}=2i (\gamma^\mu)_{\alpha\beta}\partial_\mu\, .
$$
The explicit form of the Lagrangian in terms of its components is
\be
{\cal L}=\frac{1}{2}\left( \partial_\mu\phi \,\partial^\mu\phi
+\bar\psi\, i\! \!\not\!\partial \psi +F^2 \right) + {\cal W}'(\phi)F
-\frac{1}{2}{\cal W}''(\phi)\bar\psi\psi\, .
\label{lag}
\ee
We will take ${\cal W}(\Phi )$ to be an odd function of  $\Phi$, i.e. ${\cal W}(\Phi ) $= $- {\cal W}(-\Phi )$.
In this paper we work with the following Super Polynomial Model (SPM)
\be
{\cal W}(\Phi)
=\frac{m^2}{4\,\lambda} \,\Phi - \frac{\lambda}{3}\,\Phi^3\, ,
\label{SPM}
\ee
where the parameters $m$ and $\lambda$ are  real numbers.
Thus, the Lagrangian can now be written as
\begin{align}
{\cal L} &=\frac{1}{2}\left( \partial_\mu\phi)(\,\partial^\mu\phi\right)
+\frac{1}{4}m^2\phi^2-\frac{1}{2}\lambda^2\phi^4-\frac{m^4}{32\lambda^2}
+\frac{1}{2}\bar\psi\, i\! \!\not\!\partial \psi+\lambda\phi\bar\psi\psi\, .
\label{mainlag}
\end{align}

Note that in this model the fermion is {\it ab initio} massless.
However, as we shall show in subsection \ref{ssec:continuum}, the interactions induce a dynamically generated mass at the tree level.
This mass turns out to be independent of the coupling constant $\lambda$ for $\lambda>0$,
and is equal to the mass of the elementary boson, due to SUSY.
In the following sections we investigate the solutions for $\phi$ and $\psi$ at the classical and tree level as well as their renormalization.

\section{Bosonic solutions}
\label{sec:boson}
\setcounter{equation}{0}
In this section we briefly review the properties of classical kink solution as well as its bosonic fluctuations.
In the absence of the Fermi field $\psi$, the field equation for $\phi$ derived from the Lagrangian (\ref{mainlag})
possesses static classical solitary wave solutions, including the kink which has the following form
\be
\phi_{\,0}(x) = \frac{m}{2\lambda}\,  \tanh (\frac{mx}{2})\, .
\label{kink}
\ee
The classical energy of the kink is given by
\be
-\int {\cal L}[\Phi_0]{\rm d}x=\Delta {\cal W}\equiv {\cal
W}[\phi (x=\infty )]-
{\cal W}[\phi (x=-\infty )]\, .
\label{mass}
\ee
Substituting the asymptotic values of $\phi$, i.e. $\phi(\infty)=\frac{m}{2\lambda}$ and $\phi(-\infty)=-\frac{m}{2\lambda}$,
into \mbox{Eq.\ (\ref{mass})}, we find the classical kink mass as follows
\be
M_{0}=\frac{m^3}{6\lambda^{2}}.
\ee
The eigenvalue equation for the quantum fluctuations about the kink, i.e. $\eta(x)\equiv\phi(x)-\phi_0(x)$,
can be obtained by expanding the equation for $\phi$ derived from \mbox{Eq.\ (\ref{mainlag})} and is given by
\be
\left\{-\frac{1}{2}\frac{\partial^2}{\partial z^2}+\left[3\tanh^2 (z)-1\right]\right\}\tilde\eta_n(z)
=\frac{2\omega^2_n}{m^2}\tilde\eta_n(z)\, ,
\label{B_Sch}
\ee
where $z=mx/2$ and $\tilde\eta(z)=\eta(x)$.

\begin{figure}[htp]
\centering
\includegraphics[scale=.8]{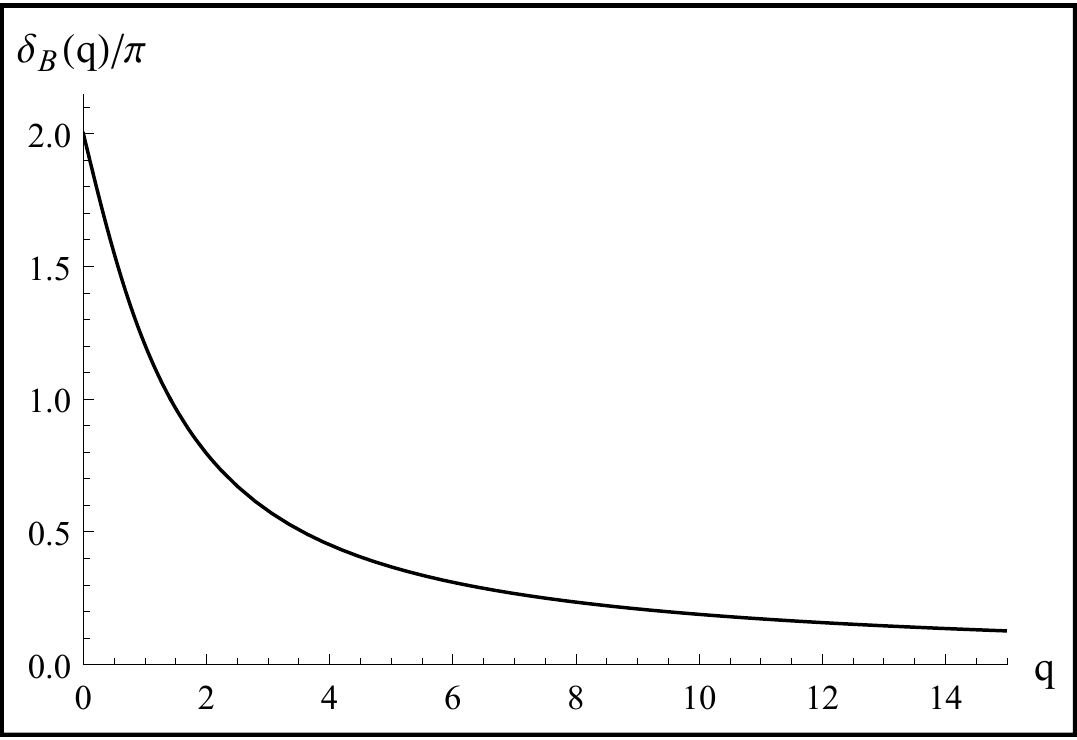}
\caption{\label{fig:Boson_Phase_Shift} \small The graphical representation of the Bosonic phase shift.}
\end{figure}

The exact solutions of \mbox{Eq.\ (\ref{B_Sch})} are well known\cite{Raj1996,Morse}.
There are two discrete levels whose energies and wavefunctions are as follows,
\begin{subequations} \label{B-modes}
\begin{align}
&\omega_{{B}_{0}}^{2}=0,\qquad \tilde\eta_{0}(z)=N_{0}\frac{1}{\cosh^2(z)}\, ,\\
&\omega_{{B}_{1}}^{2}=\frac{3}{4}m^2,\qquad \tilde\eta_{1}(z)=N_{1}\frac{\sinh (z)}{\cosh^2(z)}\, ,
\end{align}
{\textmd The continuum states are given by}
\begin{align}
\omega_{{B}_{q}}^{2}=\frac{m^2}{2}\left(\frac{1}{2}q^2+2\right), \qquad \tilde\eta_{q}(z)=\frac{{e}^{iqz}}{N(q)}\left[3\tanh^2(z)-1-q^2-3iq\tanh (z)\right] \label{contq}.
\end{align}
\end{subequations}
The asymptotic forms of $\tilde\eta_q (z)$ are given by
\begin{equation}
\lim_{z\to\pm\infty}\tilde\eta_q(z)=\exp\left[i\left(qz\pm\frac{1}{2}\delta_B (q)\right)\right],
\end{equation}
where
\begin{equation}
\delta_B (q)=-2\tan^{-1} \left(\frac{3q}{2-q^2}\right)
\end{equation}
is the bosonic phase shift of the scattering states given in \mbox{Eq.\ (\ref{contq})}, and is shown in Fig.\;\ref{fig:Boson_Phase_Shift}.
The allowed values of $q$ can be discretized by the periodic boundary condition in a box of length $L$ as follows,
\begin{align}
q_n\left(\frac{mL}{2}\right)+\delta_B\left(q_n\right)=2n\pi, \quad n\in\mathbb Z\, .
\label{B-PBC}
\end{align}
In the $L \to\infty $ limit,
\be
\sum_{q_n} \longrightarrow\frac{1}{2\pi}\int^{+\infty}_{-\infty} {\rm d}q \left[\frac{mL}{2}+\frac{\partial}{\partial q}\delta_B\left(q\right)\right]\, .
\label{bcontlimit}
\ee

\section{Fermion fluctuations}
\label{sec:fermion}
\setcounter{equation}{0}

We shall now consider the fermion fluctuations.
The Euler-Lagrange equation for the Fermi field $\psi$ derived from the Lagrangian (\ref{mainlag}) is
\be
{\left(i{\not}{\partial}+2\lambda\phi\right)\psi=0}\, .
\label{fermion_soliton}
\ee
Ignoring the back-reaction of the fermion on the kink, as it is almost invariably done, we set $\phi=\phi_0$ which is the kink.
We choose $\psi(x,t)$ to have the form
\begin{align}
\psi(x,t)=e^{-i\omega_{F} t}\left(
\begin{matrix}
\psi^{(1)}(x) \\
\psi^{(2)}(x)
\end{matrix}
\right)\, ,
\label{psi_matrix}
\end{align}
where $\omega_F$ is a real variable.
Substituting \mbox{Eq.\ (\ref{psi_matrix})} into \mbox{Eq.\ (\ref{fermion_soliton})}, one obtains
\begin{align}
{\left(\omega_F\gamma^0+i\gamma^1\partial_x+2\lambda\phi_0\right)\left(
\begin{matrix}
\psi^{(1)}(x) \\
\psi^{(2)}(x)
\end{matrix}
\right)}=0\, .
\label{psi_phi_matrix}
\end{align}
Using the representation introduced in section \ref{sec:minimal} for the $\gamma$ matrices,
\mbox{Eq.\ (\ref{psi_phi_matrix})} can be represented as follows
\be
\left(
\begin{matrix}
-\partial_x+2\lambda\phi_{0}(x) & -i\omega_{F} \\
i\omega_{F} & \partial_x+2\lambda\phi_{0}(x)
\end{matrix}
\right)
\left(
\begin{matrix}
\psi^{(1)}(x) \\
\psi^{(2)}(x)
\end{matrix}
\right)=0\, .
\label{psi_ode}
\ee
The decoupled equations are as follows
\begin{subequations}
\begin{align}
\frac{{\rm d}^2\psi^{(1)}(z)}{{\rm d} z^2}+\left\{\omega\rq{}_{F}^2-2\left[1+\tanh^2(z)\right]\right\}\psi^{(1)}(z)=0\, ,\label{secorda}\\
\frac{{\rm d}^2\psi^{(2)}(z)}{{\rm d} z^2}+\left\{\omega\rq{}_{F}^2+2\left[1-3\tanh^2(z)\right]\right\}\psi^{(2)}(z)=0\, .
\label{secordb}\end{align}
\end{subequations}
where we have rescaled the original parameters of the model as follows:
$z=mx/2$ and $\omega\rq{}_{F}=\left(2/m\right)\omega_{F}$.
Now, we parametrize our equations by defining two new parameters:
\be
\epsilon_j=\omega\rq{}_{F}^2+(-1)^{j}2\, ,\quad\textmd{and}\quad v_j=4+(-1)^{j}2\, ,
\label{epsilonv}
\ee
where $j=\{1,2\}$  denotes the upper or lower component, respectively.
Therefore, both second order differential equations \mbox{Eq.\ (\ref{secorda})} and \mbox{Eq.\ (\ref{secordb})} can be written in the same following form
\be
\frac{{\rm d}^2\psi^{(j)}(z)}{{\rm d} z^2}+\left[\epsilon_{j}-v_{j}\tanh^2(z)\right]\psi^{(j)}(z)=0\, .
\label{fermion_sch}
\ee
Our goal is to find the exact eigenvalues and eigenfunctions for this Schrodinger-like equation, analytically.
We assume the following form for $\psi^{(j)}$
\be
\psi^{(j)}(z)=\sech^b(z)F^{(j)}(z)\, .
\label{psi_suppose}
\ee
Hereafter we shall omit the superscript and subscript $j$, which refer to the component of the spinor, where no confusion could arise.
By substituting the ansatz given in \mbox{Eq.\ (\ref{psi_suppose})} into \mbox{Eq.\ (\ref{fermion_sch})}, we obtain
\begin{align}
\sech^b(z)
\bigg\{\frac{{\rm d}^2F(z)}{{\rm d}z^2}-2b\tanh(z)\frac{{\rm d}F(z)}{{\rm d}z}
+\left[\epsilon-v+b^2
+\left(v-b(b+1)\right)\sech^2(z)\right]F(z)\bigg \}=0\, .
\end{align}
Since $b$ is an arbitrary parameter, we can set
$\epsilon-v+b^2=0$.
Thus, the differential equation for $F$ is
\begin{align}
\frac{{\rm d}^2F(z)}{{\rm d}z^2}-2 b\tanh(z)\frac{{\rm d}F(z)}{{\rm d}z}+\left[v-b(b+1)\right]\sech^2(z)F(z)=0\, .
\end{align}
With the change of variables $u=\frac{1}{2}\left[1-\tanh(z)\right]$ the above equation takes on the following form
\begin{align}
u(u-1)\frac{{\rm d}^2 F(u)}{{\rm d}u^2}+\left[(2b+1+1)u-(b+1)\right]\frac{{\rm d} F(u)}{{\rm d}u}+\left[b(b+1)-v\right]F(u)=0\, .
\label{mainhyper}
\end{align}
This is the differential equation for the Gauss hypergeometric functions.
Therefore, this equation has the following exact solution
\be\label{gauss_hyper}
A~{}_{2}F_{1}(\alpha,\beta;\gamma;u)+
B~u^{1-\gamma}{}_{2}F_{1}(\alpha+1-\gamma,\beta+1-\gamma;2-\gamma;u)\, ,
\ee
where $\alpha=b+\frac{1}{2}-\sqrt{v+\frac{1}{4}}$,
$\beta=b+\frac{1}{2}+\sqrt{v+\frac{1}{4}}$ and $\gamma=b+1$.

\subsection{Fermionic bound states}
\label{ssec:bound}

In this subsection we investigate the bound states for the \mbox{Eq.\ (\ref{psi_ode})}.
Without any loss of generality, we choose $b$ to be a real positive constant.
Then the second term in \mbox{Eq.\ (\ref{gauss_hyper})} diverges as $u\to0~(z\to\infty)$.
Therefore we have to set $B=0$ for the bound states.
Then the solution which is finite at $z=\infty$ is
\begin{align}
\psi(z)&=\sech^b(z)F(z)\nonumber\\
&=\frac{N}{\left(e^{z}+e^{-z}\right)^b}
~_{2}F_{1}\left(b+\frac{1}{2}-\sqrt{v+\frac{1}{4}}
,b+\frac{1}{2}+\sqrt{v+\frac{1}{4}};1+b;\frac{e^{-z}}{e^{z}+e^{-z}}\right)\, ,
\label{wavefdiscrete}
\end{align}
where $N$ is the normalization factor.
The bound state wave function should tend to zero when $|z|\to\infty$.
The solution given in \mbox{Eq.\ (\ref{wavefdiscrete})} has the following asymptotic behavior as $z\to-\infty$,
\begin{align}
\frac{\Gamma(b+1)\Gamma(b)e^{-b z}}
{\Gamma\left(b+\frac{1}{2}-\sqrt{v+\frac{1}{4}}\right)
\Gamma\left(b+\frac{1}{2}+\sqrt{v+\frac{1}{4}}\right)}
+\frac{\Gamma(b+1)\Gamma(-b)e^{b z}}
{\Gamma\left(\frac{1}{2}-\sqrt{v+\frac{1}{4}}\right)\Gamma\left(\frac{1}{2}+\sqrt{v+\frac{1}{4}}\right)}\, .
\end{align}
The first term goes to infinity since $b>0$.
For $\psi(z)\to0$ as $z\to-\infty$, the argument of the gamma function in the denominator of the first term should be a semi-negative integer
\be
b+\frac{1}{2}-\sqrt{v+\frac{1}{4}}=-n\, ,\qquad n\in\mathbb N\cup \{0\}\, .
\label{maincond}
\ee
Since $b>0$, there is the following limit on $n$
\be                                 
0\le n<\sqrt{v+\frac{1}{4}}-\frac{1}{2}\, .
\label{limit}
\ee
Combining $\epsilon-v+b^2=0$ and \mbox{Eq.\ (\ref{maincond})}, we obtain
\be
\epsilon_n=v-\left[\sqrt{v+\frac{1}{4}}-(n+\frac{1}{2})\right]^2\, .
\label{quancond}
\ee
From this condition the permissible values of energies $\omega_{{F}_{n}}$ can be extracted.
We want to calculate the ​​allowed values of energies for the real two-component spinor $\psi(z,t)$.
For the upper component, \mbox{Eq.\ (\ref{quancond})} leads to the following bound states
\be
{\omega^{(1)}_{{F}_{0}}}^2=\frac{3}{4}m^2,\qquad\psi^{(1)}_{0}(z)=\frac{N_{1}}{2\cosh(z)}\, .
\label{minus}
\ee
Similarly, for the lower component we obtain
\begin{align}
&{\omega^{(2)}_{{F}_{0}}}^2=0\, ,~~\quad\qquad\psi^{(2)}_{0}(z)=\frac{N_{2}}{2+2\cosh\left({2 z}\right)}\, ,
\label{plus0}\\
&{\omega^{(2)}_{{F}_{1}}}^2=\frac{3}{4}m^2,\qquad\psi^{(2)}_{1}(z)=\frac{N_{2}\sinh(z)}
{2 \cosh^{2}(z)}\, .
\label{plus1}
\end{align}
The superscript of $\omega_F$ denotes component of the spinor (1 for upper and 2 for lower) and
the subscript denotes the allowed value of $n$ for that component.
The solution with $n=1$ for the spinor $\psi (z, t)$ can be written as follows:
\be
\psi_{1}(z,t)=\left(
\begin{matrix}
e^{-i\omega^{(1)}_{F_{0}}t}~~\psi^{(1)}_{0}(z)\\
e^{-i\omega^{(2)}_{F_{1}}t}~~\psi^{(2)}_{1}(z)
\end{matrix}
\right)\, .
\ee
By substituting $\psi_{1}(z,t)$ into \mbox{Eq.\ (\ref{psi_ode})}, we find $N_{2}=-i\sqrt{3}N_{1}$.
Thus,
\be
\psi_{1}(z,t)=N_{1}\frac{\sqrt{3}}{2}
e^{-i\omega_{F_{1}}t}\left(
\begin{matrix}
\frac{1}{\sqrt{3}\cosh(z)}\\
-i\frac{\sinh(z)}
{\cosh^{2}(z)}
\end{matrix}
\right)\, ,\qquad
\omega_{F_{1}}=\pm\frac{\sqrt{3}}{2}m\, .
\ee
Also from \mbox{Eq.\ (\ref{plus0})} the single non-degenerate bound state with $n=0$ and $\omega_{{F}_{0}}=0$
has the following form
(see also refs. \cite{Raj1996} and \cite{Jackiw})
\be
\psi_{0}(z,t)=N_{0}\left(
\begin{matrix}
0\\
\cosh^{-2}(z)
\end{matrix}
\right)\, ,
\ee
where $N_{0}$ is the normalization constant.

Figure \ref{fig:F_Levels} shows the bound energies of the fermion as a function of $\lambda$.
Invariance of the system under particle conjugation symmetry is evident in the figure since the bound
energies are symmetric with respect to line $\omega_{\mathrm{bound}}=0$.
As can be seen, the number of fermionic bound states  and their energies are constant for $\lambda>0$.
This is in accordance with SUSY, since the bosonic sector has the same property (see \mbox{Eqs.\ (\ref{B-modes})}).
We have previously done a spectral analysis of the Jackiw-Rebbi (JR) model which is analogous to the fermion-soliton sector
here but without SUSY\cite{Farid1,Farid2}. There the Yukawa coupling constant is a free parameter and not constrained to be related
to the coupling in the bosonic sector.
In that problem the number and energies of the bound states change as $\lambda$ increases.

\subsection{Fermionic continuum states}
\label{ssec:continuum}

\begin{figure}[htp]
\centering
\includegraphics[scale=.8]{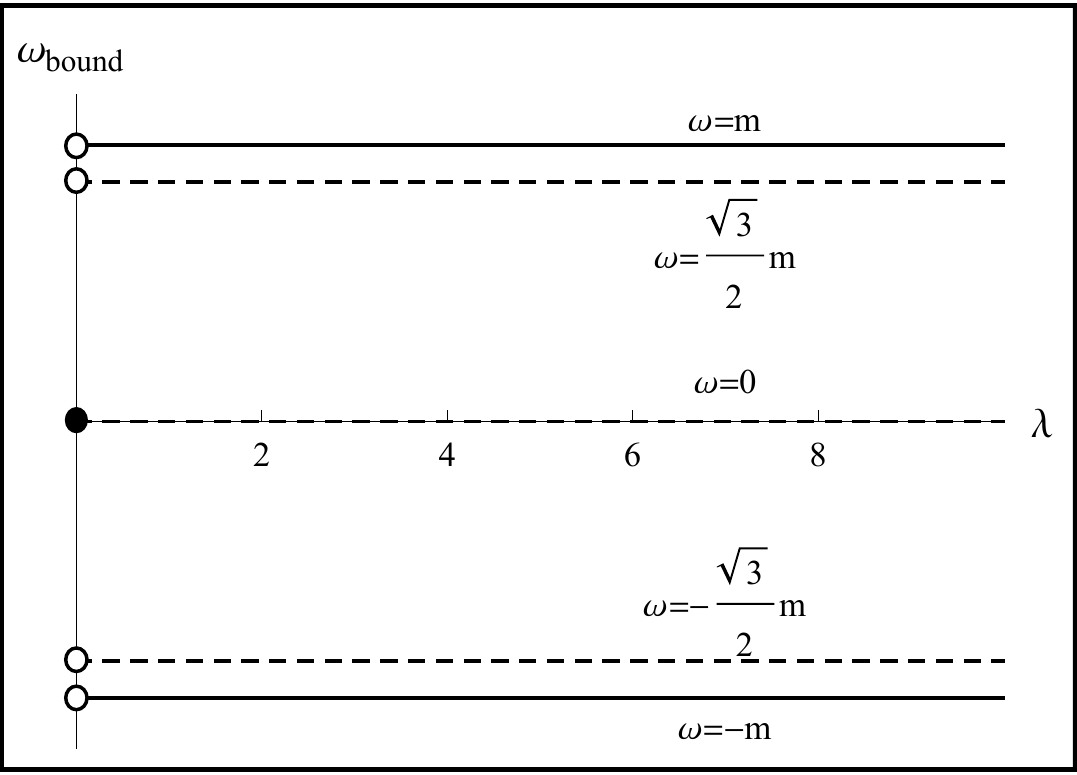}
\caption{\label{fig:F_Levels} \small Fermionic bound state energies in the prescribed soliton background field.
The graph shows the energies of the bound states (dashed lines) as a function of $\lambda$ with $\omega=0$, $\omega=\pm\frac{\sqrt{3}}{2}m$, and
the threshold ones at $\omega=\pm m$ .
Notice that \mbox{Eqs.\ (\ref{lag}, \ref{SPM})} indicate that at $\lambda=0$ the SUSY model
is in general not well defined, and if one insists on attributing a mass to the Fermi field it can only be zero.
The full and empty circles at $\lambda=0$ are meant to reflect these facts.}
\end{figure}

To obtain the fermionic continuum states, it is sufficient to set the variable $b=-ik$,
where $k$ is the wave vector of the scattering states.
As a result, we obtain
\be
\epsilon=k^{2}+v\,~~~~\textmd{or}~~~~\omega^2_{F_{k}}=\frac{m^{2}k^{2}}{4}+m^2\, .
\ee
Notice that a dynamical mass has been generated for the fermion at the tree level, which is
independent of $\lambda$, and is equal to the mass of the elementary boson field,
all valid for $\lambda>0$.
This is precisely what we expect from SUSY.
However, this dynamical mass generation has unusual intricacies associated with it which we shall now explain.
It is apparent from structure of the main Lagrangian given in \mbox{Eqs.\ (\ref{lag}, \ref{SPM})}
that the SUSY model is not well defined at $\lambda=0$.
For $\lambda>0$ the theory is well defined.
In \cite{Farid1,Farid2} we have discovered that for the JR model there is also a dynamical mass generation for the
fermion which is proportional to the coupling constant.
In this sense the dynamical mass generation in the SUSY case is discontinuous (see  Fig.\;\ref{fig:F_Levels}).

The wave function corresponding to the first independent solution given in \mbox{Eq.\ (\ref{gauss_hyper})} is
\begin{align} \label{wavefcont}
&\psi_{k}(z)=N(k){\left(e^{z}+e^{-z}\right)^{ik}}
          {}_{2}F_{1}\left(-ik+\frac{1}{2}-\sqrt{v+\frac{1}{4}}
         ,-ik+\frac{1}{2}+\sqrt{v+\frac{1}{4}};1-ik;\frac{e^{-z}}{e^{z}+e^{-z}}\right)\, .
\end{align}
By applying \mbox{Eq.\ (\ref{psi_ode})} and substituting the value of $v$ and $u$ for the upper and lower components we obtain
\begin{align}
\psi_{k}(z)=N(k){\left(e^{z}+e^{-z}\right)^{ik}}\left(
\begin{matrix}
{}_{2}F_{1}\left(-1-ik,2-ik;1-ik;\frac{e^{-z}}{e^{z}+e^{-z}}\right)\\
-\frac{\left(k+2i\right)}{\sqrt{4+k^2}}~{}_{2}F_{1}\left(-2-ik,3-ik;1-ik;\frac{e^{-z}}{e^{z}+e^{-z}}\right)
\end{matrix}
\right)\, ,
\label{spectra_cont}
\end{align}
where $N(k)$ is a normalization factor.
Ignoring the extra constant multiplicative factor of the lower component shown in \mbox{Eq.\ (\ref{spectra_cont})}, both components have the following asymptotic behavior near $z= \pm\infty$:
\begin{align}
&\lim_{z\to\infty}\psi_{k}(z)=N(k)e^{ikz}\, ,\nonumber\\
&\lim_{z\to -\infty}\psi_{k}(z)=\frac{N(k)\Gamma\left(1-ik\right)\Gamma(-ik){e^{ikz}}}
{{\Gamma(-ik+\frac{1}{2}-\sqrt{v+\frac{1}{4}})}
\Gamma(-ik+\frac{1}{2}+\sqrt{v+\frac{1}{4}})}\nonumber\\
&+\frac{N(k)\Gamma(1-ik)\Gamma(ik)e^{-ikz}}
{\Gamma(\frac{1}{2}-\sqrt{v+\frac{1}{4}})\Gamma(\frac{1}{2}+\sqrt{v+\frac{1}{4}})}\, .
\label{limitpsi_cont}
\end{align}
It is apparent that this is a scattering state with an incident wave from the left.
Analogous analysis shows that the second independent solution given in \mbox{Eq.\ (\ref{gauss_hyper})}
corresponds to a scattering state with an incident wave from the right.

\subsection{Phase shift}
\label{ssec:Phase}

\begin{figure}[htp]
\centering
\includegraphics[scale=.8]{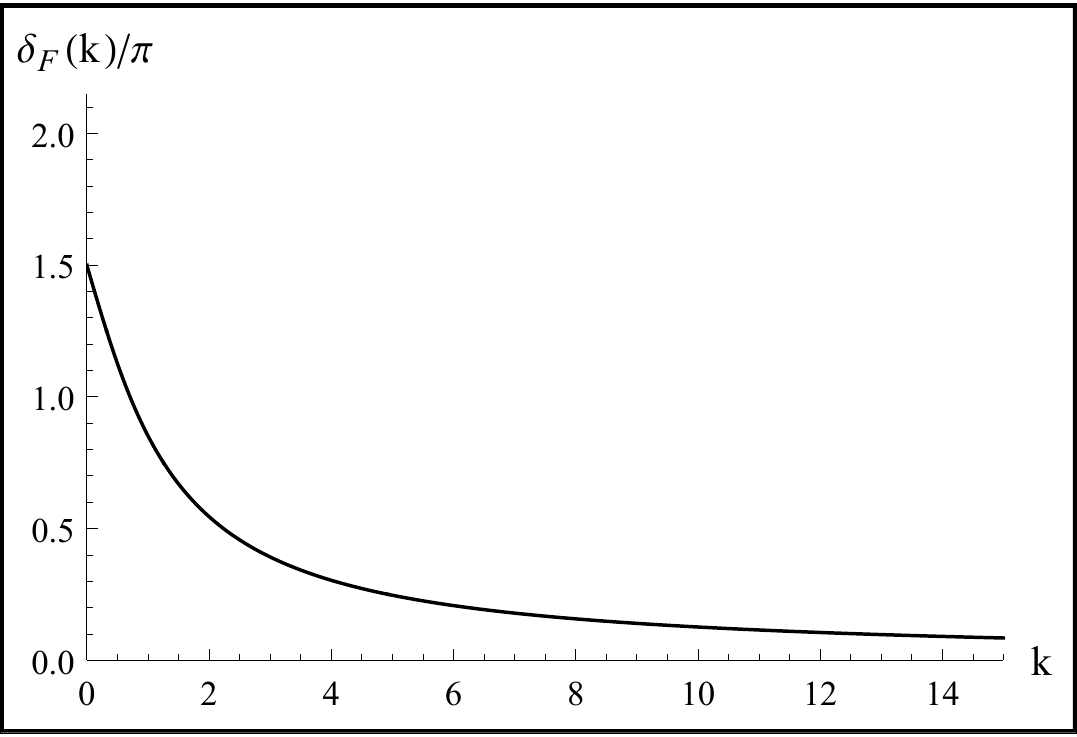}
\caption{\label{fig:Fermion_Phase_Shift} \small The graphical representation of the Fermionic phase shift.}
\end{figure}

Our calculation of the one-loop quantum correction to the SUSY kink mass includes finding the phase shifts of the scattering states,
which we shall do without any approximations.
Using \mbox{Eq.\ (\ref{limitpsi_cont})}, one can obtain the exact phase shift of the fermionic scattering states as follows,
\begin{align}
\exp\left[{i\delta_{F}(k)}\right]&= \frac{\Gamma\left(-ik+\frac{1}{2}-\sqrt{v+\frac{1}{4}}\right)\Gamma\left(-ik+\frac{1}{2}+\sqrt{v+\frac{1}{4}}\right)}
{{\Gamma(1-ik)}\Gamma(-ik)}\nonumber\\
&=\frac{\beta\left(-ik+\frac{1}{2}-\sqrt{v+\frac{1}{4}},-ik+\frac{1}{2}+\sqrt{v+\frac{1}{4}}\right)}{\beta\left(1-ik,-ik\right)}\, ,
\end{align}
where $\delta_{F}(k)$ is the fermionic phase shift of the scattering states of \mbox{Eq.\ (\ref{limitpsi_cont})}.
For the upper component $\psi^{(1)}_{k}(z) $, using \mbox{Eq.\ (\ref{epsilonv})}, we obtain
\be
\delta^{(1)}_{F}(k)=\tan^{-1}\left(\frac{2k}{-1+k{^2}}\right).
\ee
Similarly we can compute the phase shift for the lower component $\psi^{(2)}_{k}(z)$ which yields
\be
\delta^{(2)}_{F}(k)=\tan^{-1}\left[\frac{6(-2+k^{2})k}{4-13k{^2}+k^{4}}\right].
\ee
We have defined the fermionic phase shift to be the average of its
values for the upper and lower components and this definition has been checked
using the strong and weak forms of the Levinson theorem\cite{Gousheh2}.
This definition yields a unique expression for the phase shift for the
spinor $\psi$ and has been used and checked in several different models,
all giving consistent results\cite{Gousheh2,Gousheh3}.
As we shall shortly show, it also yields consistent results for this problem.
\begin{align}
\delta_{F}(k)&=\frac{1}{2}\left[\delta^{(1)}_{F}(k)+\delta^{(2)}_{F}(k)\right]\nonumber\\
&=\frac{1}{2}\left[\tan^{-1}\left(\frac{2k}{-1+k{^2}}\right)+\tan^{-1}\left(\frac{6(-2+k^{2})k}{4-13k{^2}+k^{4}}\right)\right]\nonumber\\
&=\frac{1}{2}\tan^{-1}\left[\frac{4k\left(5-11k^2+2k^4\right)}{-4+41k{^2}-26k^4+k^6}\right]\, ,
\end{align}
the result is shown in Fig.\;\ref{fig:Fermion_Phase_Shift}.
The allowed value of $k$ can be discretized by the anti-periodic boundary condition in a box of length $L$ as follows,
\be
k_n\left(\frac{mL}{2}\right)+\delta_F\left(k_n\right)=(2n+1)\pi, \quad n\in\mathbb Z
\label{F-PBC}\, .
\ee
In the $L \to\infty $ limit,
\be
\sum_{k_n} \longrightarrow\frac{1}{2\pi}\int^{+\infty}_{-\infty} {\rm d}k \left[\frac{mL}{2}+\frac{\partial}{\partial k}\delta_F\left(k\right)\right]\, .
\label{fcontlimit}
\ee

\noindent In the next subsection we investigate the validity of this phase shift
using both the strong and weak forms of the Levinson theorem.

\subsection{Levinson theorem}
\label{ssec:Levinson}

In this subsection we study the properties of the fermionic phase shift in connection with both the weak and strong forms of the of the
Levinson theorem. With this study we not only check the consistency of our results, but also gain further insight into the spectral
properties of the solutions. By imposing a boundary on the system and then sending the boundary
to infinity, or by the more formal S-matrix arguments, one can find the following 
relation (see for example \cite{Graham_Thesis}),
\be
\rho(k)-\rho_{0}(k)=\frac{1}{\pi}\frac{{\rm d}\delta(k)}{{\rm d}k}\, ,
\ee
which relates the difference between the density of the continuum
states in the free and interacting cases, i.e. the spectral deficiency in that continuum, to the phase shift.
This relationship holds for each of the positive and negative continua, separately, and hence for the sum of the two.
This leads to the weak form of the Levinson theorem for the Dirac equation which can be written
in the following form (see for example\cite{Gousheh2})
\be
\Delta\delta_{F}(k)\equiv\Big[\delta^{{\rm sky}}_{F}(0)-\delta^{{\rm sky}}_{F}(\infty)\Big]
+\Big[\delta^{{\rm sea}}_{F}(0)-\delta^{{\rm sea}}_{F}(\infty)\Big]=\left(N+\frac{N_{t}}{2}-\frac{N_{t}^{0}}{2}\right)\pi\, ,
\label{Dirac_Levinson}
\ee
where $N$ is the total number of bound states,
$N_{t}$ the total number of the threshold bound states at the given strength of the potential,
and $N_{t}^{0}$ the number of bound states at zero strength of the potential, i.e. the free Dirac case.
The threshold bound states count as half bound states and should be included as shown in \mbox{Eq.\ (\ref{Dirac_Levinson})}
only in one spatial dimension or for the S-states.

Now we discuss the Levinson theorem for the fermionic phase shifts.
First we should note that the system possesses particle conjugation symmetry which mandates that the spectrum
be completely symmetric about the line $E=0$. This in turn implies that $\delta(k=\infty)=0$, which is consistent
with the phase shift shown in Fig.\;\ref{fig:Fermion_Phase_Shift}.
The fermions in the prescribed soliton background field have bound states
at $\omega=0$, $\omega=\pm(\sqrt{3}/{2})m$, and threshold bound states at $\omega=\pm m$
(see  Fig.\;\ref{fig:F_Levels}).
At $\lambda=0$, the fermion is massless and two threshold bound states
at $\omega=0$ separate the positive and negative energy continua (see  Fig.\;\ref{fig:F_Levels})
and, as in JR model, these states join to form a nondegenerate zero-energy bound state\cite{Farid1}.
Therefore for $\lambda>0$, $3/2$ states have exited either one of the continua.
According to the strong form of the Levinson theorem\cite{Gousheh2}, the phase shift
at $k=0$ for either one of the continua should be $3\pi/2$, and Fig.\;\ref{fig:Fermion_Phase_Shift} confirms this.
Now, the weak form of the Levinson theorem as expressed in \mbox{Eq.\ (\ref{Dirac_Levinson})}
is easily verified as follows
\be
\left(\frac{3\pi}{2}-0\right)+\left(\frac{3\pi}{2}-0\right)=
\left[3+2\left(\frac{1}{2}\right)-2\left(\frac{1}{2}\right)\right]\pi=3\pi\,  .
\ee

\section{The one-loop correction to the SUSY kink mass}
\label{sec:one-loop}
\setcounter{equation}{0}

In this section we compute the radiative correction to the SUSY kink mass.
The kink mass including the one-loop correction due to both bosonic and fermionic fluctuations, is given by
\cite{Kaul,Litvintsev},
\begin{align}
M_{1}=M_0+\frac{1}{2}\left(\sum{\omega_{B}}-\sum{\omega_{F}}\right)+\Delta m_{\textmd{ct}}\, ,
\end{align}
 where $M_0$ is the classical energy of the kink or the kink mass in the zero order, the second term
 is related to the Casimir energy contribution, and the last term is the
 required counterterm to remove the divergence of the second term.
 Obviously the bound states\rq{} contribution to the Casimir energy is zero due to SUSY.
 Using the results obtained in the previous sections, the above equation can be rewritten as follows
\begin{align}
M_{1}&=\frac{m^{3}}{6\lambda^{2}}+\frac{1}{2}\left[\sum_n\left({\frac{m}{\sqrt{2}}\sqrt{2+\frac{q^2_{n}}{2}}-\frac{m}{\sqrt{2}}\sqrt{2+\frac{k^2_{n}}{2}}}\right)\right]
+\Delta m_{\textmd{ct}}\, .
\label{main_m_1}
\end{align}

Consider a box of length $L$.
Using \mbox{Eq.\ (\ref{B-PBC})} and \mbox{Eq.\ (\ref{F-PBC})},
$k_ {n}$ and $q_ {n}$ can be related to each other as follows 
\begin{align}
2n\pi=k_n\left(\frac{mL}{2}\right)+\delta_F\left(k_n\right)-\pi=q_n\left(\frac{mL}{2}\right)+\delta_B\left(q_n\right)\, .
\end{align}
Thus, the term in brackets in \mbox{Eq.\ (\ref{main_m_1})} can be simplified as follows:
\begin{align}
&\sum_n\left({\frac{m}{\sqrt{2}}\sqrt{2+\frac{q^2_{n}}{2}}-\frac{m}{\sqrt{2}}\sqrt{2+\frac{k^2_{n}}{2}}}\right)\nonumber\\
&=\sum_n\frac{-k_{n}\left[-\delta_F(k_{n})+\delta_B(q_{n})+\pi\right]}{L}\left(1+\frac{k^{2}_{n}}{4}\right)^{-\frac{1}{2}}+{\cal O}\left(\frac{1}{L^2}\right).
\end{align}
Therefore, in the $L\to\infty$ limit, the total mass of the kink soliton up to one-loop correction is
\begin{align}
M_{1}&=\frac{m^{3}}{6\lambda^{2}}-\frac{1}{4\pi}\int^{+\infty}_{-\infty} {\rm d}k\left\{\frac {k\delta(k)}{L\sqrt{4+k^2}}
\left[\frac{mL}{2}+\frac{\partial}{\partial k}\delta_{F}\left(k\right)\right]\right\}
+\Delta m_{\textmd{ct}}\nonumber\\
&=\frac{m^{3}}{6\lambda^{2}}-\frac{m}{8\pi}\int^{+\infty}_{-\infty} {\rm d}k\frac {k\delta(k)}{\sqrt{4+k^2}}\nonumber\\
&-\frac{1}{4\pi L}\int^{+\infty}_{-\infty} {\rm d}k\left[\frac {k\delta(k)}{\sqrt{4+k^2}}\frac{\partial}{\partial k}\delta_{F}\left(k\right)\right]
+\Delta m_{\textmd{ct}},
\end{align}
where $\delta(k)\equiv-\delta_{F}(k)+\delta_{B}(k)+\pi $ and $\delta_{B}(k)=\delta_{B}(q)+{\cal O}\left(1/L\right)$.
We can eliminate the ${\cal O}(1/L)$ terms and obtain,
\begin{align}
M_1=\frac{m^{3}}{6\lambda^{2}}-\frac{m}{8\pi}\int^{+\infty}_{-\infty} {\rm d}k\frac {k\delta(k)}{\sqrt{4+k^2}}+\Delta m_{\textmd{ct}}\, .
\end{align}
Integrating by parts, the corrected SUSY kink mass becomes
\begin{align}
M_1=\frac{m^{3}}{6\lambda^{2}}-\frac{m}{8\pi}\left[\delta(k)\sqrt{4+k^2}\right]^{+\infty}_{-\infty}+\frac{m}{8\pi}\int^{+\infty}_{-\infty} {\rm d}k\sqrt{4+k^2}\frac{\rm {d}}{\rm {d}k}\delta(k)
+\Delta m_{\textmd{ct}}\, .
\end{align}
The total phase shift, including both fermionic and bosonic contribution, is
\begin{align}
\delta(k)=-\frac{1}{2}\tan^{-1}\left[\frac{4k\left(5-11k^2+2k^4\right)}{-4+41k{^2}-26k^4+k^6}\right]
-2\tan^{-1} \left(\frac{3k}{2-k^2}\right)+\pi\, .
\end{align}
After some algebra, it is straightforward to find the following expression for the SUSY kink mass in the one-loop order 
\begin{align}
M_1&=\frac{m^{3}}{6\lambda^{2}}-\frac{m}{2\pi}-\frac{m}{4\pi}\int^{+\infty}_{-\infty}\frac{{\rm d}k}{\sqrt{1+k^2}}+\Delta m_{\textmd{ct}}\, .
\label{LogDivergent}
\end{align}
Note that the integral is logarithmically divergent.
This divergence is exactly canceled by the SUSY counterterm.
The mass counterterm $\delta m$ is fixed by requiring that the tadpole counterterm cancels the tadpole graphs\cite{Litvintsev2000,Peskin}:
\be
\raisebox{-0.6cm}{\includegraphics[width=7.mm]{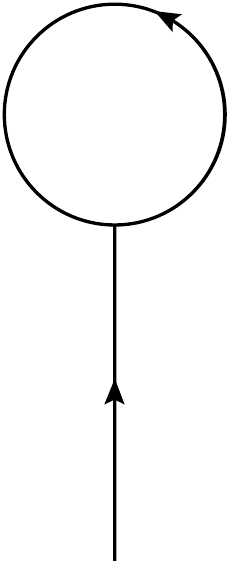}}\hspace{2mm}
+\hspace{2mm}\raisebox{-0.6cm}{\includegraphics[width=7.mm]{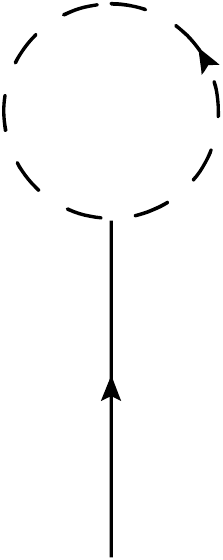}}\hspace{2mm}
+\hspace{2mm}\raisebox{-0.6cm}{\includegraphics[width=3.mm]{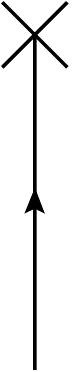}}\hspace{2mm}=0\, .
\ee
Consequently,
\begin{align}
&3im\lambda\int_{-m\Lambda}^{m\Lambda}\frac{{\rm d}^2k}{(2\pi)^2}\frac{i}{k^2-m^2}
-i\lambda\int_{-m\Lambda}^{m\Lambda}\frac{{\rm d}^2k}{(2\pi)^2}\frac{i\Tr\left(\! \!\not\!k+m\right)}{k^2-m^2}-\frac{im}{2\lambda}\delta m=0\, ,
\end{align}
where $m\Lambda$ is a momentum cut-off.
After some algebra, we obtain the SUSY counterterm
\begin{align}
\delta m=\frac{\lambda^2}{2 \pi} \int_{-m\Lambda}^{m\Lambda}\frac{{\rm d}k}{\sqrt{m^2+k^2}}\, .
\end{align}
The bosonic loop (solid circle) contribution to the SUSY counterterm is proportional
to $\frac{3 \lambda^2}{2 \pi}$ and the fermionic loop 
(dotted circle) is proportional to $-\frac{\lambda^2}{\pi}$.
Hence, we obtain the following result for the last term in \mbox{Eq.\ (\ref{LogDivergent})}
\begin{align}
\Delta m_{\textmd{ct}}&=\frac{m}{2\lambda^2}\delta m=\frac{m}{4 \pi} \int_{-\Lambda}^{\Lambda}
\frac{{\rm d}k}{\sqrt{1+k^2}}\, .
\label{susyct}
\end{align}
Therefore, we obtain
\begin{align}
M_1=\frac{m^{3}}{6\lambda^{2}}-\frac{m}{2\pi}\, .
\end{align}
Our result is in complete agreement with the accepted value of the supersymmetric kink mass reported previously.

\section{Conclusions}
\label{sec:conclusion}
\setcounter{equation}{0}

In this paper we calculate the radiative correction to the kink mass up to one-loop
in a (1+1)-dimensional theory with minimal supersymmetry.
Our method is based on the exact calculation of the spectra and the phase shifts, including both the
bosonic and fermionic contributions, which enables us
to compute the one-loop correction to the SUSY kink mass.
The exact solutions in the bosonic sector has already been available.
We solve the equation of the fermion in the presence of the kink exactly and analytically and find the whole spectrum
of the fermionic fluctuations including both the bound and continuum states.
We find that while the Fermi field is {\it ab initio} massless, contrary to the elementary Bose field,
a dynamical mass is generated at the tree level which is exactly equal to that of the
elementary Bose field and is independent of the coupling constant due to SUSY.
Also, the number of bound states of the fermion is independent of the parameters of the model,
including the parameters of kink and this happens just because of SUSY.
Moreover, the fermionic spectrum is completely independent of the coupling $\lambda$,
which causes a unique Casimir energy and vacuum polarization for all values of $\lambda$.
The presence of the kink causes a complication in the exact calculation of the fermionic phase shift.
Our prescription to define the fermionic phase shift is to take the average of its
values for the upper and lower components.
This enables us to compute a unique and exact expression for the fermionic phase shift.
We also check the consistency of the resulting fermionic phase shift by the Levinson theorem.
We then explicitly compute the Casimir energy by the use of the phase shift method.
We use the renormalized perturbation theory to cancel the
divergent part of the SUSY kink mass by the appropriate SUSY counterterm.
We show that this procedure is sufficient to obtain the accepted
value for the mass of the kink in a (1+1)-dimensional theory
with minimal SUSY at the one-loop level, and there is no need
to use other theorems, such as the Levinson theorem, to remedy
the divergences or even fix the finite part.

\section*{Acknowledgments}
\label{Acknowledgments}
\addcontentsline{toc}{section}{\numberline{}Acknowledgments}

SMH and FC would like to thank Noah Graham for
helpful conversations, suggestions and references and also
Leila Shahkarami for useful discussions.

\addcontentsline{toc}{section}{\numberline{}References}

\end{document}